\documentclass[aps,pra,twocolumn,nofootinbib, superscriptaddress,10pt]{revtex4-1}

\usepackage{mathdots}
\usepackage{amsfonts,amssymb,amsmath}
\usepackage[]{graphics,graphicx,epsfig}
\usepackage{amsthm}
\usepackage{csquotes}
\MakeOuterQuote{"}
\usepackage{epstopdf}

\def\identity{\leavevmode\hbox{\small1\kern-3.8pt\normalsize1}}

\newtheorem{lemma}{Lemma}

\newcommand{\ket}[1]{\left | #1 \right\rangle}

\newcommand{\rank}{\operatorname{rank}}
\newcommand{\supp}{\operatorname{supp}}
\newcommand{\Tr}{\operatorname{Tr}}
\newcommand{\tr}{\operatorname{tr}}

\newcommand{\NOT}{\mathrm{NOT}}

\newcommand{\pro}{\mathrm{pro}}
\newcommand{\rmi}{\mathrm{i}}
\newcommand{\rme}{\operatorname{e}}
\newcommand{\rmW}{\mathrm{W}}
\newcommand{\5}{\operatorname{\uppercase\expandafter{\romannumeral5}}}

\renewcommand{\epsilon}{\varepsilon}

\def\eqref#1{\textup{(\ref{#1})}}
\newcommand{\eref}[1]{Eq.~\textup{(\ref{#1})}}
\newcommand{\lref}[1]{Lemma~\ref{#1}}

\def\<{\langle}  
\def\>{\rangle}  

\begin{document}

\title{Implementation of generalized measurements on a qudit via quantum walks}

\author{Zihao Li}
\affiliation{Department of Physics and Center for Field Theory and Particle Physics, Fudan University, Shanghai 200433, China}
\affiliation{State Key Laboratory of Surface Physics, Fudan University, Shanghai 200433, China}

\author{Haoyu Zhang}
\affiliation{Department of Physics and Center for Field Theory and Particle Physics, Fudan University, Shanghai 200433, China}
\affiliation{State Key Laboratory of Surface Physics, Fudan University, Shanghai 200433, China}

\author{Huangjun Zhu}
\email{zhuhuangjun@fudan.edu.cn}
\affiliation{Department of Physics and Center for Field Theory and Particle Physics, Fudan University, Shanghai 200433, China}
\affiliation{State Key Laboratory of Surface Physics, Fudan University, Shanghai 200433, China}
\affiliation{Institute for Nanoelectronic Devices and Quantum Computing, Fudan University, Shanghai 200433, China}
\affiliation{Collaborative Innovation Center of Advanced Microstructures, Nanjing 210093, China}

\begin{abstract}
Quantum measurements play a fundamental role in quantum mechanics and quantum information processing, but it is not easy to implement generalized measurements, the most powerful  measurements allowed by quantum mechanics.  Here we propose a simple recipe for implementing generalized measurements on a qudit via quantum walks.  With this recipe,	any discrete quantum measurement  can be implemented via a one-dimensional discrete quantum walk; the number of steps is only two times the number of measurement outcomes. As an illustration, we present a unified  solution for implementing arbitrary symmetric informationally complete  measurements in dimension 3.
\end{abstract}

\date{\today}
\maketitle

\section{Introduction}
Quantum measurements are the key for extracting information from quantum systems and are  indispensable in many quantum information processing protocols, such as quantum tomography, metrology, and communication \cite{Paris04, Haya05, Lvovsky09, Giov11, NielC00}.  The most general and powerful  measurements in quantum mechanics are known as positive operator-valued measures (POVMs) \cite{NielC00}.  Mathematically,  a POVM is composed  of a collection $\{E_i \}$ of
positive operators, known as POVM elements, which sum up to the identity, namely,  $\sum_i E_i=\openone$.  If the POVM is performed on a quantum state $\rho$, then  the probability of obtaining outcome $i$ is given by the Born rule, $p(i) = \Tr (\rho E_i)$.

POVMs are useful to  quantum information processing because they can achieve many tasks, such as unambiguous quantum state discrimination \cite{Barn09,Bae15} and optimal quantum tomography \cite{MassP95, ZhuH18, Hou18}, which cannot be achieved by traditional projective measurements (also known as von Neumann measurements). In addition, POVMs can be constructed that are informationally complete (IC), so they can extract all required information in one go.  Prominent examples of IC POVMs include symmetric informationally complete POVMs (SIC POVMs) \cite{Zaun11, ReneBSC04, Fuchs17}, which are of interest in many research areas, including quantum tomography \cite{Scot06,ZhuE11,ZhuH18} and foundational studies \cite{FuchS13,ApplFSZ17,Zhu16Q}.

Theoretically, any POVM can be implemented by first applying a unitary transformation on a joint system composed of the original system and an ancilla of sufficiently large Hilbert space, and then performing a projective measurement on the joint system.
However, the unitary transformation to be applied is usually quite complicated and the dimension of the subspace on which it has nontrivial action increases steadily with the number of outcomes (or POVM elements).
Therefore, it is usually quite daunting in practice to implement a POVM  in this "textbook" way by brute force.
In addition, no  simple  recipe is known for implementing  general POVMs on a qudit that is amenable to experiments. So far only  several simple POVMs have been realized in the lab \cite{Ling06, Clarke01, Mosley06, Tabia12, Meden11, Waldherr12}.
Most known implementations  are based on \emph{ad hoc} methods tailored to particular situations.

Quantum walks are another important tool  in various branches of quantum information science, including quantum computation  and quantum simulation \cite{Magn11, Child09, Love10, Nico14}. In particular, quantum walks on certain sparse graphs can realize universal quantum computation \cite{Child09}. Our interest in quantum walks is stimulated by their applications
 in quantum measurements. Recently,  Kurzy\'nski and  W\'ojcik (KW) \cite{KurzW13} proposed a method for implementing general POVMs on a qubit via one-dimensional discrete quantum walks. Note that this result does not follow from Ref.~\cite{Child09} because here the walker is restricted to a one-dimensional chain instead of sparse graphs. Shortly, the KW scheme  was  demonstrated successfully in photonic quantum systems \cite{Bian15, Zhao15}. Unfortunately,  no general recipe is known so far for implementing qudit POVMs via quantum walks, although  a special two-qubit  POVM was realized  \cite{Hou18}. Here the situation is much more complicated because it is already  nontrivial to establish a sensible connection between the states of the coin and the movements of the walker.
Also, it is much more difficult to devise suitable coin operators due to the large Hilbert space. A breakthrough on this problem is of interest not only to theoretical studies but also to experimental quantum information processing.

In this paper we propose a simple but general recipe for implementing qudit POVMs via quantum walks. We show that 	any discrete  POVM on a qudit  can be implemented via a one-dimensional discrete quantum walk. For a rank-1 POVM, the number of steps required is only two times the number of POVM elements, that is, the number of measurement outcomes. Compared with the traditional approach, the size of the unitary transformation to be applied can be  reduced significantly. To implement a  SIC~POVM in dimension $d$,  for example, the traditional approach requires the application of a unitary transformation of size at least $d^2$ (typically on the order of $d^3$). With our approach, by contrast, this size can be reduced to $d$. To demonstrate the power of this recipe, we present a unified  solution for implementing all SIC~POVMs in dimension~3. Our study is instrumental to exploring the physics in higher-dimensional Hilbert spaces and to achieving various quantum information-processing tasks that rely on generalized measurements. Meanwhile, it is of intrinsic interest to the foundational studies on quantum measurements.

\section{Implementation of POVMs via quantum walks}
A discrete-time quantum walk is a process in which the movement of a particle (walker) on a one-dimensional lattice is controlled by its  internal  state (coin).
The state $|x\rangle\otimes|c\rangle$ of the joint system is determined  by two indices, where $x=\dots, -1,0,1,\dots$ represents the walker position, and $c=0,1,\dots,d-1$ labels the coin state.
Here $d$ is the dimension of the coin Hilbert space, and $\{|c\>\}_{c=0}^{d-1}$ forms an orthonormal basis. Each step of the quantum walk corresponds to a unitary operator that has  the form $U(t)=TC(t)$, where $C(t)=\sum_x|x\rangle\langle x|\otimes C(x,t)$, with $C(x,t)$ being position-dependent coin operators and $T$ is the conditional translation operator. In the case of a qubit ($d=2$), the operator $T$ usually takes on the form
\begin{equation}
T=\sum_x \bigl(|x+1,0\rangle\langle x,0|+|x-1,1\rangle\langle x,1|\bigr).
\end{equation}
Measurement of the walker position after certain steps of the quantum walk effectively implements a POVM on the coin state.  What is not so obvious is that any POVM on a qubit can be realized in this way by choosing the coin operators properly \cite{KurzW13}.

Little is known about implementing  qudit POVMs   via quantum walks despite the significance of this problem.
First, there is no canonical choice for the translation operator, without which  we would be stuck at the beginning. Second, it is much more difficult to devise suitable coin operators because the characteristics of a qubit usually do not generalize to a qudit.  To break this deadlock requires some inspiration.

\begin{figure*}
	\center{\includegraphics[scale=0.78]{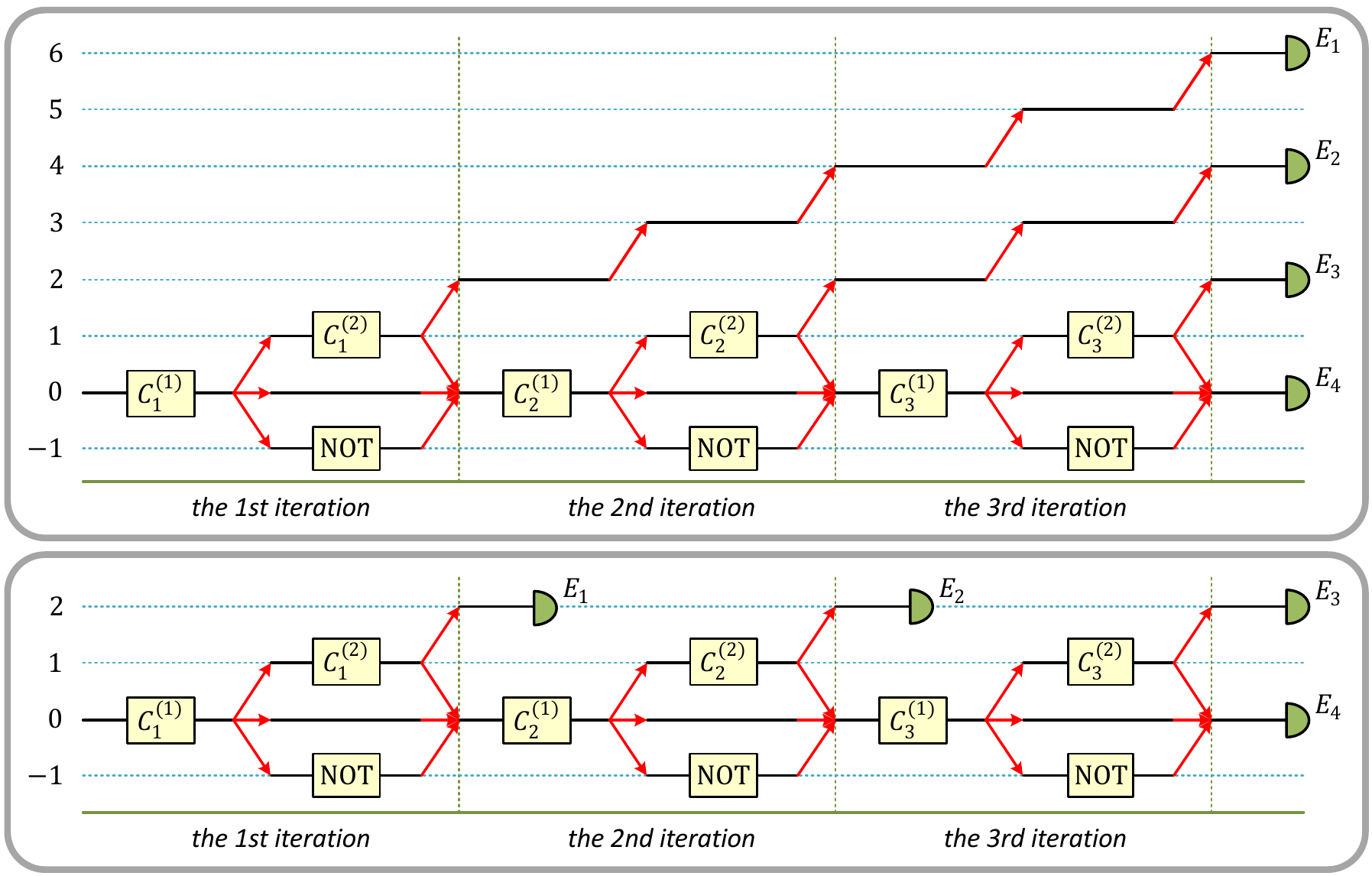}}
	\vspace{-2mm}
	\caption{\label{fig:QWPOVM} Schematic diagrams for   implementing  a rank-1 POVM on a qudit via a discrete quantum walk.
	Here the target POVM $\{E_1, E_2, E_3, E_4\}$ has four elements and is  realized  using a three-iteration (six-step) quantum walk.
The initial coin state at position $x=0$ is  the qudit state of interest. The upper plot illustrates the algorithm presented in Sec.~\ref{sec:QWalgorithm},
and  the four detectors $E_1$ to $E_4$ correspond to the four POVM elements, respectively.	
The detector $E_1$ ($E_2$) can also be put at position~2 after the 1st (2nd) iteration, as illustrated in the  lower plot, because no coin operators are present along the corresponding paths subsequently.	
	}
\end{figure*}

Here we propose the following translation operator
\begin{equation}\label{eq:Translation}
T=\sum_x \Big(|x+1,0\rangle\langle x,0|+|x-1,1\rangle\langle x,1|+\sum_{j=2}^{d-1}|x,j\rangle\langle x,j|\Big),
\end{equation}
which means  the walker moves up (down) if the coin state is $|0\>$ ($|1\>$), but stands still otherwise (cf.~Fig.~\ref{fig:QWPOVM}). What is surprising is that this simple translation operator enables us to  implement arbitrary qudit POVMs as long as the coin operators are chosen wisely.

\section{Example}
Before presenting the general recipe, let us first consider  a special  qutrit POVM that  is composed of four  elements of the form $E_i = \frac{3}{4}(|\psi_i\rangle\langle\psi_i|)$, where
\begin{equation}\label{eq:Tetrahedron}
|\psi_i\rangle = \frac{1}{\sqrt{3}}\sum_{j=0}^{2}(-1)^{\delta_{i,j+2}}|j\>,\quad i=1,2,3,4.
\end{equation}

Now we introduce a six-step quantum-walk protocol that is able to perform this POVM on a  qutrit. The joint system is initialized at the state $|0\>\otimes |\varphi\>$, where the coin state $|\varphi\>$ corresponds to the state to be measured.
In  the first step, all coin operators are trivial (equal to the identity) except for the one corresponding to position~0, which has the form
\begin{equation}
C(0,1) = \begin{pmatrix}
\frac{1}{\sqrt{3}}&\frac{1}{\sqrt{3}}&\frac{1}{\sqrt{3}}\\\frac{1}{\sqrt{6}}&\frac{1}{\sqrt{6}}&\frac{-2}{\sqrt{6}}\\\frac{1}{\sqrt{2}}&\frac{-1}{\sqrt{2}}&0
\end{pmatrix}.
\end{equation}
In the second step, the nontrivial  coin operators read
\begin{equation}
C(-1,2) = \begin{pmatrix}0&1&0\\1&0&0\\0&0&1\end{pmatrix}, \;\; C(1,2) = \begin{pmatrix}\frac{\sqrt{3}}{2}&\frac{1}{2}&0\\\frac{1}{2}&\frac{-\sqrt{3}}{2}&0\\0&0&1\end{pmatrix}.
\end{equation}
In the third step, the  only  nontrivial coin operator reads
\begin{equation}
C(0,3) = \begin{pmatrix}
\frac{-1}{\sqrt{6}}&\frac{1}{\sqrt{3}}&\frac{-1}{\sqrt{2}}\\\frac{-1}{\sqrt{6}}&\frac{1}{\sqrt{3}}&\frac{1}{\sqrt{2}}\\\sqrt{\frac{2}{3}}&\frac{1}{\sqrt{3}}&0
\end{pmatrix}.
\end{equation}
In the last three steps, the nontrivial coin operators are $C(-1,4)$ and $C(-1,6)$, which are identical to $C(-1,2)$.
After six steps, measurement of  the walker position implements the desired POVM. More precisely, the four POVM elements $E_1, E_2, E_3$, and $E_4$ correspond to  the four positions  $x=6,4,2$, and $0$, respectively.

To confirm our claim, note that the probability $p_k$  of finding the particle at position $k$ is given by
\begin{equation}
p_k =\tr [U(|0\>\<0|\otimes |\varphi\>\<\varphi|)U^\dag (|k\>\<k|\otimes\openone )]=\tr(\Omega_k |\varphi\>\<\varphi|),
\end{equation}
where   $U=TC(6)TC(5)\cdots TC(1)$ is the unitary operator generated by the coin operators and translation operator, and   $\Omega_k$ is the POVM element corresponding to  position $x=k$.  Simple calculation shows that
\begin{equation}\label{eq:Omega}
\Omega_k=\Tr_{\rmW}\{(|0\>\<0|\otimes\openone)U^{\dagger} \left(|k\rangle\langle k| \otimes \openone \right) U\},
\end{equation}
where "$\Tr_{\rmW}$" denotes the partial trace over the walker degree of freedom.  Now it is straightforward to verify that
\begin{equation}
\Omega_6=E_1, \quad \Omega_4=E_2, \quad \Omega_2=E_3, \quad \Omega_0 =E_4,
\end{equation}
so the POVM specified in \eref{eq:Tetrahedron} is indeed realized by the above quantum-walk protocol.
Note that the program can also be terminated after the third step, because the desired POVM  can already be realized at this point.
More precisely, the  four positions  $x=3,1,-1$, and $0$ after the third step
correspond to the four POVM elements $E_1, E_2, E_3$,  and $E_4$,  respectively.

\section{A simple algorithm for realizing arbitrary POVMs}
Now we are ready to propose a general algorithm for realizing arbitrary discrete POVMs. To this end, it suffices to focus on rank-1 POVMs because higher-rank POVM elements can be expressed as sums of rank-1 elements.

\subsection{\label{sec:QWalgorithm}Algorithm}
Let $\{E_1, E_2,\dots,E_n\}$ be a rank-1 POVM, where each POVM element $E_i$ is proportional to the projector onto a pure state, that is, $E_i = a_i|\psi_i\rangle\langle\psi_i|$, with $0<a_i\leq 1$. The following algorithm implements this POVM with only $2(n-1)$ steps, as illustrated in the upper plot in Fig.~\ref{fig:QWPOVM}; see the Appendix for a variant algorithm.
\begin{enumerate}
\item Initialize the quantum walk at position $x=0$ with the coin state corresponding to the qudit state to be measured; set $i = 1$.
\item For $i=1,2,\ldots, n-1$,
\begin{enumerate}
\item Apply coin operator $C_i^{(1)}$ at position $x=0$ and identity elsewhere, and then apply translation operator $T$.

\item Apply coin operator $C_i^{(2)}$ at position $x=1$, $\NOT$ at position $x=-1$, and identity elsewhere, and then apply translation operator $T$.

\item $i=i +1$.
\end{enumerate}
\item Measure the walker position, then  the positions $2n-2, 2n-4, \dots, 0$ correspond to the POVM elements $E_1, E_2, \dots, E_n$, respectively.
\end{enumerate}

Here the coin operators $\NOT$ and $C_i^{(2)}$ have  the forms
\begin{equation}\label{eq:C(2)}
\NOT = \begin{pmatrix}\ 0&1&\\\ 1&0&\\&&\openone_{d-2}\end{pmatrix}, \quad  C_i^{(2)} = \begin{pmatrix}\alpha'_i&\beta'_i&\\\beta'_i&-\alpha'_i&\\&&\openone_{d-2}\end{pmatrix}.
\end{equation}
The parameters $\alpha'_i, \beta'_i$ and the coin operator $C_i^{(1)}$ depend on the POVM and can be determined recursively.
More precisely, $C_i^{(1)}$ is chosen such that
\begin{equation}\label{eq:C^(1)}
C_{i}^{(1)\dagger}|0\rangle=b_i^{-1}K_{i-1}^{\dag+}|\psi_{i}\rangle,
\end{equation}
where $b_i:=\|K_{i-1}^{\dag+}|\psi_{i}\rangle\|$,  $K_0=\openone$,
\begin{equation}
K_l=\bigg(|0\>\<1|+\beta'_l|1\>\<0|+\sum_{k=2}^{d-1}|k\>\<k|\bigg)C_l^{(1)}K_{l-1} \quad (l\geq1),
\end{equation}
and $K_{i-1}^{\dag+}$ is the Moore-Penrose generalized inverse of $K_{i-1}^\dagger$ \cite{Penrose55, Bern05}, which reduces to the usual inverse when $K_{i-1}^\dag$ is invertible. The parameters
$\alpha'_i$ and $\beta'_i$ read
\begin{equation}\label{eq:C^(2)}
\alpha'_{i} = \sqrt{a_i}b_i,\quad  \beta'_{i}=\sqrt{1-\alpha'^2_{i}}.
\end{equation}
If $\alpha'_{i}=1$, then the coin operator $C_i^{(2)}$ can also be replaced by the identity operator.

In the above algorithm the detector $E_j$ at position $2(n-j)$ after $n-1$ iterations can also be put at position~2 after the $j$th iteration for $j=1,\ldots, n-2$, as illustrated in the  lower plot in Fig.~\ref{fig:QWPOVM}, because subsequently no coin operators are present along the corresponding paths.
By virtue of the above algorithm any rank-1 discrete POVM with $n$ elements on a qudit can be realized with $(n-1)$ iterations, that is, $2(n-1)$ steps. In general, each POVM element can be decomposed into a sum of at most $d$ rank-1 POVM elements, so a general POVM with $n$ elements can be realized with at most $2(n d-1)$ steps.

\subsection{Proof of universality}
To verify that the above algorithm indeed implements the desired POVM, let us consider the evolution of the joint state of the walker and coin. Suppose $|\varphi_0\rangle=|\varphi\>$ is the qudit state to be measured, then the initial state of the joint system is $|0\>\otimes|\varphi_0\rangle$.
After  applying  the coin operator $C_1^{(1)}$ and  translation operator $T$,  the system state evolves into
\begin{equation}
\alpha_1|1,0\rangle+\beta_1|-1,1\rangle+|0\rangle\otimes|\varphi'_0\rangle,
\end{equation}
where $\alpha_1=\langle0|C_1^{(1)}|\varphi_0\rangle$, $\beta_1=\langle1|C_1^{(1)}|\varphi_0\rangle$, and the ket  $|\varphi'_0\rangle=\sum_{k=2}^{d-1}|k\rangle\langle k|C_1^{(1)}|\varphi_0\rangle$  is not  normalized.

Next, we turn to step 2(b) ($i=1$). At position $x=-1$ the coin state changes from $|1\rangle$ to $|0\rangle$, so the walker has no chance to go down beyond $x=-1$. At position $x=1$ coin operator $C_1^{(2)}$ is applied. After applying the translation operator $T$, the joint state reads
\begin{align}
&\alpha'_1\alpha_1|2,0\rangle+\beta_1|0,0\rangle+\beta'_1\alpha_1|0,1\rangle+|0\rangle\otimes|\varphi'_0\rangle\nonumber\\
&=\alpha'_1\alpha_1|2,0\rangle+|0\rangle\otimes|\varphi_1\rangle,
\end{align}
where $\alpha'_1=\langle0|C_1^{(2)}|0\rangle$, $\beta'_1=\langle1|C_1^{(2)}|0\rangle$, and the ket  $|\varphi_1\rangle = K_1|\varphi_0\rangle$ is not necessarily normalized.

Similarly, the joint state after the $i$th iteration reads
\begin{equation}\label{eq:JointStatej}
|0\rangle\otimes|\varphi_i\rangle+\sum_{l=1}^{i}\alpha'_{i-l+1}\alpha_{i-l+1}|2l,0\rangle,
\end{equation}
where we have $|\varphi_l\rangle=K_l|\varphi_0\rangle$, $\alpha_l=\langle0|C_l^{(1)}|\varphi_{l-1}\rangle$, and $\beta_l=\langle 1|C_l^{(1)}|\varphi_{l-1}\rangle$.

If we measure the walker position now, then the probabilities of finding the particle at positions $x=0$ and $x=2(i-j+1)$ ($1\leq j\leq i$) read
\begin{align}
&p_{x=0}=\tr(|\varphi_i\rangle\langle\varphi_i|)=\tr\big(K_i^\dagger K_i|\varphi_0\rangle\langle\varphi_0|\big),\\
&p_{x=2(i-j+1)}=|\alpha'_j\alpha_j|^2\nonumber \\
&=|\alpha'_j|^2\tr\big[\big(K_{j-1}^\dagger C_j^{(1)\dagger} |0\rangle\langle0| C_j^{(1)} K_{j-1}\big)(|\varphi_0\rangle\langle\varphi_0|)\big].
\end{align}
So the position $x=0$ corresponds to the  POVM element $\Omega_0=K_i^\dagger K_i$, and the  position $x=2(i-j+1)$ corresponds to the  POVM element
\begin{align}\label{eq:Pi}
\Pi_j&=|\alpha'_j|^2\bigl(K_{j-1}^\dagger C_j^{(1)\dagger} |0\rangle\langle0| C_j^{(1)} K_{j-1}\bigr)\nonumber\\
&=a_j  K_{j-1}^\dagger K_{j-1}^{\dag+}|\psi_{j}\rangle\langle\psi_{j}|K_{j-1}^{+}K_{j-1}\nonumber\\
&=a_j P_{K_{j-1}}|\psi_{j}\rangle\langle\psi_{j}|P_{K_{j-1}}=a_j|\psi_{j}\rangle\langle\psi_{j}|=E_j,
\end{align}
where $P_{K_{j-1}}$ is the projector onto  $\supp(K_{j-1})$. Here  the second equality follows from Eqs.~\eqref{eq:C^(1)}  and \eqref{eq:C^(2)} as long as $b_j>0$; the condition  $b_j>0$ and  the fourth
equality follow from the fact
$|\psi_{j}\rangle\in\supp(K_{j-1}^{\dag +} )=\supp(K_{j-1})$,  which can be proved by induction. When $j=1$, the condition $|\psi_{j}\rangle\in\supp( K_{j-1})$ holds because $K_0=\openone$.
In general, the condition  follows from the inequality below
\begin{equation}\label{eq:2}
E_{j}= a_j|\psi_{j}\rangle\langle\psi_{j}|\leq \openone -\sum_{l=1}^{j-1}E_l=K_{j-1}^\dagger K_{j-1}
\end{equation}
given that $\Pi_l=E_l$ for $l=1, 2,\ldots, j-1$ by the induction hypothesis and that $K_{j-1}^\dagger K_{j-1}=\openone-\sum_{l=1}^{j-1}\Pi_l$.
Meanwhile, the above equation implies that
\begin{equation}\label{eq:alpha'}
\alpha_{j}'^2 =a_{j}b_{j}^2=
a_{j}\bigl\|K_{j-1}^{\dag+}|\psi_{j}\rangle\langle\psi_{j}|K_{j-1}^+\bigr\|\leq 1,
\end{equation}
so the coin operator $C_{j}^{(2)}$ is well defined.

We emphasize that the POVM element  $\Pi_j$ in \eref{eq:Pi} does not depend on the iteration number $i$ as long as $1\leq j\leq i$.
By the end of the  $(n-1)$th iteration, the joint state is given by \eref{eq:JointStatej} with $i=n-1$. Measurement of the walker position then realizes the POVM elements $E_1, E_2, \dots, E_n$, which correspond to finding the particle at positions $2n-2, 2n-4, \dots, 0$, respectively.
Therefore, any rank-1 discrete POVM with $n$ elements  can be realized by the above algorithm with $(n-1)$ iterations, that is, $2(n-1)$ steps.

Actually, the quantum-walk protocol for implementing the POVM specified in \eref{eq:Tetrahedron} follows from the general algorithm presented above.
In this case, calculation shows that $\alpha'_2=\alpha'_3=1$ and $K_{2}^{\dag+}|\psi_{3}\rangle\propto|0\rangle$, so we have
$C_{2}^{(2)}=C_{3}^{(1)}=C_{3}^{(2)}=\openone$.

\subsection{Generation of post-measurement states}
Our quantum-walk protocol can also be adapted to generate desired post-measurement quantum states, say, $|\zeta_l\>$ for $l=1,2,\ldots,n$. If we apply
the  coin operator $C_{n}^{(1)}$ [defined according to \eref{eq:C^(1)}] at position $x=0$ before the measurement, then the joint state would evolve into
\begin{equation}\label{eq:JointStateA}
\alpha_{n}|0,0\>+\sum_{l=1}^{n-1}\alpha'_{n-l}\alpha_{n-l}|2l,0\>,
\end{equation}
where $\alpha_n=\langle0|C_n^{(1)}|\varphi_{n-1}\rangle$. Equation~\eqref{eq:JointStateA} follows from
\eref{eq:JointStatej} and the fact
that  $C_{n}^{(1)}|\varphi_{n-1}\> \propto|0\>$. To see this fact, note that $K_{n-1}^\dag K_{n-1}=E_n$, so that $K_{n-1}$ has rank~1, $K_{n-1}^{\dag+}\propto K_{n-1}$, and
\begin{equation} |\varphi_{n-1}\>=K_{n-1}|\varphi_0\>\propto K_{n-1}^{\dag+} |\psi_n\>\propto C_{n}^{(1)\dag}|0\>.
\end{equation}
To obtain desired post-measurement states, now it suffices to apply a coin operator at position $2l$ to turn $|0\>$ into $|\zeta_{n-l}\>$ for $l=0, 1, \ldots, n-1$. In this way, our protocol can also be applied to realizing quantum channels that are based on measurements and preparation.

\section{Implementation of qutrit SIC POVMs}\label{app:discordandentanglement}
In a $d$-dimensional Hilbert space, a SIC POVM \cite{Zaun11, ReneBSC04, Fuchs17} is composed of $d^2$ subnormalized projectors onto pure states
$E_i = |\psi_i\rangle\langle\psi_i|/d$, with equal pairwise fidelity,
\begin{equation}
|\langle\psi_i|\psi_j\rangle|^2=\frac{d\delta_{ij}+1}{d+1}, \quad  i,j=1,2,\dots,d^2.
\end{equation}
Such POVMs are of great interest in
quantum information processing and quantum foundations \cite{Fuchs17}.
For a  qutrit,  all  SIC POVMs are covariant with respect to the Heisenberg-Weyl group \cite{ReneBSC04, Zaun11, Appl05, Zhu10, Hugh16, Szol14}, which is generated by the shift operator $X$ and phase operator $Z$,
\begin{equation}
X = \begin{pmatrix} 0&0&1 \\ 1&0&0 \\ 0&1&0\end{pmatrix}, \ \ Z = \begin{pmatrix} \ 1&0&0 \\ \ 0&\omega&0 \\ \ 0&0&\omega^2\end{pmatrix}, \ \ \omega=\rme^{\rmi\frac{2\pi}{3}}.
\end{equation}
Up to a unitary transformation, any SIC POVM in dimension 3 can be generated by the Heisenberg-Weyl group from a fiducial state of the form
\begin{equation}
	|\psi\rangle = \frac{1}{\sqrt{2}}(|1\>-\rme^{\rmi t}|2\>), \quad t\in [0,2\pi).
\end{equation}
Given the fiducial state, all  states of the SIC POVM can be expressed as follows,
\begin{equation}
|\psi_i\rangle = X^j Z^k|\psi\rangle, \quad j,k=0,1,2;\quad  i=3j+k+1.
\end{equation}

By virtue of the general algorithm presented above, here we offer a unified solution for implementing all SIC POVMs in dimension 3 via discrete quantum walks.
In total, 8 iterations and 16 steps are required to implement each qutrit SIC POVM. The coin operators $C_i^{(1)}$ featuring in the algorithm for $i=1,2,\ldots, 8$  are presented below,
\begin{widetext}
\begin{align}
&C_1^{(1)} = \begin{pmatrix} 0 &\frac{1}{\sqrt{2}}&-\frac{\rme^{-\rmi t}}{\sqrt{2}} \\ 1&0&0 \\ 0&\frac{1}{\sqrt{2}}&\frac{\rme^{-\rmi t}}{\sqrt{2}}\end{pmatrix},&~
&C_2^{(1)} = \begin{pmatrix} 0&-\frac{1}{\sqrt{3}} & -\rmi\sqrt{\frac{{2}}{3}} \\1&0&0 \\ 0 &\sqrt{\frac{{2}}{3}}&-\frac{\rmi}{\sqrt{3}} \end{pmatrix},&~
&C_3^{(1)} = \begin{pmatrix} 0&-\frac{1}{\sqrt{6}}&-\sqrt{\frac{5}{6}}  \\ 1&0&0 \\ 0&\sqrt{\frac{5}{6}}&-\frac{1}{\sqrt{6}} \end{pmatrix},&\nonumber \\
&C_4^{(1)} = \begin{pmatrix} -\frac{\rme^{-\rmi t}}{\sqrt{3}} & \frac{\rme^{\rmi t}}{q^2\sqrt{3}} & \frac{q\rme^{\rmi t}}{\sqrt{3}}\\0& \frac{1}{\sqrt{2}}&-\frac{\rmi}{\sqrt{2}}\\\frac{2q^2\rme^{-2\rmi t}}{\sqrt{6}}&\frac{1}{\sqrt{6}}&\frac{\rmi}{\sqrt{6}} \end{pmatrix},&~
&C_5^{(1)} = \begin{pmatrix} 0&\frac{1}{q^5\sqrt{2}}&-\frac{q\rme^{\rmi t}}{\sqrt{2}} \\ 1&0&0 \\ 0&\frac{1}{\sqrt{2}}&-\frac{\rme^{\rmi t}}{\sqrt{2}} \end{pmatrix},&~
&C_6^{(1)} = \begin{pmatrix} 0&-{\frac{q^2}{\sqrt{2}}}&-\frac{1}{\sqrt{2}}\\1&0&0&\\0&\frac{1}{\sqrt{2}}&-{\frac{1}{q^2\sqrt{2}}} \end{pmatrix}, \\
&C_7^{(1)} = \begin{pmatrix} \frac{q^2\rme^{-\rmi t}}{\sqrt{2}}&0&\frac{q\rme^{\rmi t}}{\sqrt{2}} \\ 0&1&0 \\ \frac{1}{\sqrt{2}}&0&\frac{q^5\rme^{2\rmi t}}{\sqrt{2}}\end{pmatrix},&~
&C_8^{(1)} = \begin{pmatrix} 0&-\frac{1}{\sqrt{2}}&\frac{q^5\rme^{-\rmi t}}{\sqrt{2}} \\ 1&0&0 \\ 0&\frac{1}{\sqrt{2}}&\frac{q^5\rme^{-\rmi t}}{\sqrt{2}}  \end{pmatrix},& \nonumber
\end{align}
where  $q := \rme^{\rmi\frac{\pi}{6}}$.   The coin operators $C_i^{(2)}$ for $i=1,2,\ldots, 8$
are determined by the parameters $\alpha_i'$ and $\beta_i'$ presented below,
\begin{align}
&\alpha'_1=\sqrt{\frac{1}{3}}, \quad \beta'_1=\sqrt{\frac{2}{3}},&~~
&\alpha'_2=\sqrt{\frac{3}{8}}, \quad \beta'_2=\sqrt{\frac{5}{8}},&~~
&\alpha'_3=\sqrt{\frac{2}{5}}, \quad \beta'_3=\sqrt{\frac{3}{5}},&\nonumber \\~~~~~
&\alpha'_4=\sqrt{\frac{1}{2}}, \quad \beta'_4=\sqrt{\frac{1}{2}},&~~
&\alpha'_5=\sqrt{\frac{2}{3}}, \quad \beta'_5=\sqrt{\frac{1}{3}},&~~
&\alpha'_6=1, \quad \beta'_6=0,&\\
&\alpha'_7=\sqrt{\frac{2}{3}}, \quad \beta'_7=\sqrt{\frac{1}{3}},&~~
&\alpha'_8=1, \quad \beta'_8=0.&\nonumber
\end{align}
\end{widetext}
To be concrete, they are specified in \eref{eq:C(2)} when $\alpha'_i>0$ and set to the identity otherwise.
Surprisingly, all coin operators $C_i^{(2)}$ are independent of the parameter $t$ that characterizes the qutrit SIC POVM. By contrast, five out of the eight coin operators $C_i^{(1)}$ depend on the parameter $t$. Compared with previous proposals \cite{Meden11, Tabia12, Kalev12}, our solution is appealing because it follows from a universal recipe.

To verify that our protocol indeed implements the desired qutrit SIC POVM, let us consider the evolution of the joint state of the walker and coin. The  initial state of the joint system can be expressed as follows,
\begin{equation}
|\Phi_0\rangle=|x=0\rangle\otimes|\varphi\rangle=a|0,0\rangle+b|0,1\rangle+c|0,2\rangle,
\end{equation}
where $|\varphi\rangle$ is a general qutrit pure state,
\begin{equation}
|\varphi\rangle=a|0\rangle+b|1\rangle+c|2\rangle, \qquad |a|^2 + |b|^2 + |c|^2 = 1.
\end{equation}
After the first iteration, the joint state evolves  into
\begin{widetext}
\begin{equation}
|\Phi_1\rangle=\frac{b-c\rme^{-\rmi t}}{\sqrt{6}}|2,0\rangle+a|0,0\rangle+\frac{b-c\rme^{-\rmi t}}{\sqrt{3}}|0,1\rangle+\frac{b+c\rme^{-\rmi t}}{\sqrt{2}}|0,2\rangle.
\end{equation}
By the same token, after eight iterations, the joint state reads
\begin{align}
|\Phi_8\rangle&={\frac{1}{\sqrt{6}}}({b-c\rme^{-\rmi t}})|16,0\rangle+{\frac{q^4}{\sqrt{6}}}({bq^4-c \rme^{-\rmi t}})|14,0\rangle+{\frac{q^{-4}}{\sqrt{6}}}({bq^{-4}-c\rme^{-\rmi t}})|12,0\rangle\nonumber \\
&\quad +{\frac{1}{\sqrt{6}}}({c-a\rme^{-\rmi t}})|10,0\rangle+ {\frac{q^4}{\sqrt{6}}}({cq^4-a \rme^{-\rmi t}})|8,0\rangle+{\frac{q^{-4}}{\sqrt{6}}}({cq^{-4}-a\rme^{-\rmi t}})|6,0\rangle\ \nonumber \\
&\quad+{\frac{1}{\sqrt{6}}}({a-b\rme^{-\rmi t}})|4,0\rangle+{\frac{q^4}{\sqrt{6}}}({aq^4-b\rme^{-\rmi t}})|2,0\rangle -{\frac{q^{-4}}{\sqrt{6}}}({aq^{-4}-b\rme^{-\rmi t}})|0,2\rangle.
\end{align}
\end{widetext}

If we measure the walker position at this point, then the  probabilities of finding the particle at  the positions $x =16, 14, \dots, 0$ are respectively given by
\begin{equation}
\begin{aligned}
p_{x=16} & = \frac{1}{6}|{b-c\rme^{-\rmi t}}|^2 = \langle \varphi|E_1|\varphi\rangle , \\
p_{x=14} & =  \frac{1}{6}|{bq^{4}-c\rme^{-\rmi t}}|^2 = \langle \varphi|E_2|\varphi\rangle , \\
&\dots  \\
p_{x=0} & =  \frac{1}{6}|{aq^{-4}-b\rme^{-\rmi t}}|^2 = \langle \varphi|E_9|\varphi\rangle .
\end{aligned}
\end{equation}
Therefore, the above protocol can indeed  realize the desired SIC POVM.

\section{Summary}
We proposed a simple  recipe for implementing arbitrary discrete qudit POVMs via one-dimensional quantum walks.
By virtue of this recipe,  any rank-1 POVM with $n$ POVM elements can be realized via a quantum walk with $2(n-1)$ steps, and a general POVM can be realized with at most $2(nd-1)$ steps. Compared with the traditional approach, our recipe can significantly reduce the size of the unitary transformation to be applied. Notably, the size of each coin operator featuring in our algorithm does not depend on the number of POVM elements. This merit
is appealing from both theoretical and experimental perspectives.
As far as we know, this recipe is the most versatile approach for implementing general POVMs that is amenable to experiments.
For illustration, we devised an explicit protocol for implementing arbitrary  SIC~POVMs in dimension~3.
Our study not only offers valuable insight on implementing quantum measurements, but also provides a useful tool for  achieving many quantum information-processing tasks that rely on generalized measurements. In the case of a qubit, implementation of POVMs via quantum walks has been realized using photonic systems in which the photon polarization serves as the coin qubit \cite{Bian15, Zhao15}. By employing the orbital angular momentum of a photon \cite{Devlin17,Padg17}, there is a possibility of realizing our protocol in photonic systems as well. We hope that our work can stimulate further progress along this direction.

\section*{Acknowledgments}
We are grateful to Erkka Haapasalo for stimulating discussions on quantum channels. HZ is grateful to Zhibo Hou and Jiangwei Shang for stimulating discussions on quantum walks and to Christopher Fuchs for valuable comments.
This work is supported by the National Natural Science Foundation of China (Grant No. 11875110).

\appendix
\section{Alternative algorithm for implementing qudit POVMs via quantum walks}

In this appendix, we present a variant algorithm for implementing  general discrete POVMs on a qudit via  discrete quantum walks.
Here the translation operator is still given by \eref{eq:Translation}, but the idea for choosing the coin operators is different, as explained as follows. The two variants may offer complementary perspectives on implementing POVMs via quantum walks.

\subsection{Train of thoughts}
As in the main text, here we focus on rank-1 POVMs because higher-rank POVM elements can be expressed as sums of rank-1 elements.
Given  a rank-1 POVM element $E\propto|\psi\rangle\langle\psi|$, any state $|\psi^{\perp}\rangle$ that is orthogonal to $|\psi\rangle$ satisfies  $\langle \psi^{\perp}|E|\psi^{\perp}\rangle=0$ and thus can never lead to the corresponding outcome. Suppose  after some steps of the quantum walk, measurement at position $x$ corresponds to the element $E$, while the initial coin state of the quantum walk were $|\psi^{\perp}\rangle$; then the probability of finding the particle at position $x$ would be zero.

The above idea can be applied to  generate a POVM element that is proportional to $|\psi\rangle\langle\psi|$. As in the main text  the quantum walk is initialized at position $x=0$  with the coin state corresponding to the qudit state to be measured. Let  $\{|\psi_{\perp m}\rangle\} \ (m=1,2,\dots,d-1)$ be a set of orthonormal kets that are  orthogonal to $|\psi\rangle$.
If at some stage of the quantum walk, a proper unitary coin operator $C_{\pro}$ is applied at position $x$ such that, after the following translation, the probability of detecting the particle at position $x+1$ is zero for each initial state of the coin in the set $\{|\psi_{\perp m}\rangle\}_{m=1}^{d-1}$,  then position $x+1$ would correspond to a POVM element of the form $a|\psi\rangle\langle\psi|$ with  $0\leq a\leq 1$. Here we assume that the walker at position $x+1$ can only come from position $x$, but not from position $x+2$; this assumption will be guaranteed by a suitable design of the algorithm.

The task of finding the operator $C_{\pro}$ is not difficult. Suppose $|\phi_{0}\rangle, |\phi_{1}\rangle, \dots, |\phi_{d-1}\rangle$ are the coin states at position $x=0$ just before $C_{\pro}$ is applied if the initial coin states were $|\psi\rangle, |\psi_{\perp1}\rangle, \dots, |\psi_{\perp d-1}\rangle$, respectively.
It is not difficult to find a coin operator that can  transform $|\phi_{1}\rangle, \dots,|\phi_{d-1}\rangle$ into some states that are all orthogonal to $|0\rangle$. If we apply this operator as $C_{\pro}$ at position $x$, then the following translation will not move the walker to position $x+1$.

\subsection{Algorithm}

Now we are ready to propose a variant of the general algorithm for realizing an arbitrary rank-1 POVM $\{E_1, E_2,\dots,E_n\}$, where each POVM element $E_i$ is proportional to the projector onto a pure state, that is, $E_i = a_i|\psi_i\rangle\langle\psi_i|$, with $0<a_i\leq 1$.
\begin{enumerate}
\item Initialize the quantum walk at position $x=0$ with the coin state corresponding to the qudit state one wants to measure; set $i = 1$ and $j = 0$.
\item While $i < n$ do:
\begin{enumerate}
\item Apply coin operator $C_{i,j}^{1}$ at position $x=0$ and identity elsewhere, and then apply translation operator $T$.
\item Apply coin operator $C_{i,j}^{2}$ at position $x=1$, $\NOT$ at position $x=-1$, and identity elsewhere, and then apply translation operator $T$.
\item If $j=d-1$, then apply coin operator $\NOT$ at position $x=0$.
\item If $\cos\theta_i=1$, then $j=j+1$ and  apply coin operator $C_{i,j}^{3}$ at position $x=0$.
\item $i=i +1$.
\end{enumerate}
\item Measure the walker position, then the positions $2n-2, 2n-4, \dots, 0$ correspond to the POVM elements $E_1, E_2, \dots, E_n$, respectively.
\end{enumerate}

The coin operators $C_{i,j}^{1}$, $C_{i,j}^{2}$, and $C_{i,j}^{3}$ have the forms
\begin{equation}
\begin{aligned}
C_{i,j}^{1} &= |0\rangle\langle \eta_{i}|+|1\rangle\langle \eta_{i\perp1}|+\dots+|d-1\rangle\langle \eta_{i\perp d-1}|,\\
C_{i,j}^{2} &= \begin{pmatrix}
\cos\theta_i&\sin\theta_i&\\\sin\theta_i&-\cos\theta_i&\\&&\openone_{d-2}
\end{pmatrix}, \qquad 0\leq \theta_i\leq \pi/2,\\
C_{i,j}^{3} &= \openone+\!|1\rangle\langle d-j|+\!|d-j\rangle\langle1|-\!|1\rangle\langle1|-\!|d-j\rangle\langle d-j|.
\end{aligned}
\end{equation}
As shown later, $j$ always satisfies the inequality $j\leq d-1$, so  coin operators $C_{i,j}^{3}$ are well defined. The kets featuring in $C_{i,j}^{1}$ are chosen as follows. Let
$\{|\psi_{i\perp m}\rangle\}_{m=1}^{d-1}$ be an arbitrary but given set of orthonormal kets that are orthogonal to $|\psi_i\rangle$; let
$|\phi_{i0}\rangle, |\phi_{i1}\rangle, \dots, |\phi_{i,d-1}\rangle$ be the unnormalized  coin states at position $x=0$ at the beginning of the $i$th iteration when the initial coin states were $|\psi_i\rangle, |\psi_{i\perp1}\rangle, \dots, |\psi_{i\perp d-1}\rangle$, respectively.
$|\eta_{i}\rangle$ is a normalized ket that is orthogonal to $|\phi_{i1}\rangle, \dots, |\phi_{i,d-1}\rangle$, and $\{|\eta_{i\perp m}\rangle\}_{m=1}^{d-1}$ is a set of  orthonormal  kets that are orthogonal to $|\eta_i\rangle$. The form of the coin operator $C_{i,j}^{1}$ guarantees that, after $C_{i,j}^{1}$ and the subsequent translation operator are applied, the POVM element corresponding to position $x=1$ is proportional to $|\psi_i\rangle\langle\psi_i|$. The role of $C_{i,j}^{2}$ is to adjust  the magnitude of this  POVM element so that, after step 2(b),  position $x=2$ corresponds to the desired POVM element $E_i$.

When $j\geq 1$, we will prove later that  all the kets
$|\phi_{i1}\rangle, |\phi_{i2}\rangle, \dots, |\phi_{i,d-1}\>$ are orthogonal to $\ket{d-k}$  for $k=1,2,\ldots,j$  and that the dimension of $\mathrm{span}\left\{|\phi_{i1}\rangle, \dots, |\phi_{i,d-1}\rangle\right\}$ is less than $d-j$.
Therefore, we can choose the ket $|\eta_{i}\rangle$  in the orthogonal complement of the  $\mathrm{span}\{|d-1\>, |d-2\>, \dots, |d-j\>\}$ and set
$|\eta_{i\perp m}\rangle=|m\>$ for $m=d-1,d-2,\dots,d-j$.

A POVM is simple if no two POVM elements are proportional to each other. In this case $j$ can reach the value of $d-1$ only after step 2(d) when  $i=n-1$, so step 2(c) in the above algorithm is not necessary. This step is introduced so that the algorithm can  be applied to non-simple POVMs as well. For a non-simple POVM, after $j$ reaches the value of $d-1$,  application of $C_{i,d-1}^1$ is redundant because $C_{i,d-1}^1$ are  equal to the identity; in addition, step 2(d) is redundant because the condition $\cos\theta_i=1$ can never be satisfied when $j=d-1$.

\subsection{Proof of universality}

The above algorithm guarantees that the walker going up beyond $x=1$ can not move back (which clarifies our previous assumption that the walker cannot move from  $x+2$ to $x+1$). On the other hand, the coin operator $\NOT$ guarantees that the walker has no chance to go down beyond position $x=-1$. We have already shown that the operator $C_{i,j}^{1}$ helps generate a POVM element proportional to $|\psi_i\rangle\langle\psi_i|$. It remains to prove that the algorithm is able to adjust the magnitude of this element to the desired value.
We shall prove this fact by induction.

First,  consider the first iteration ($i=1$) of the subroutine 2(a)-2(e). The walker starts at position $x=0$  with  the coin state corresponding to the state to be measured. Note that $|\eta_1\>=\rme^{\rmi\xi}|\psi_1\>$, so after step 2(a)  the POVM element corresponding to position $x=1$  is $|\psi_1\rangle\langle\psi_1|$. After step 2(b) with the choice $\cos\theta_1=\sqrt{a_1}$,   position $x=2$ corresponds to the POVM element  $\cos^2\theta_1|\psi_1\rangle\langle \psi_1|=E_1$, and  position $x=0$ corresponds to the  element $\openone -E_1$.
In addition, if $a_1<1$, then $\rank(\openone -E_1)=d$, and we still have $j=0$  at the beginning of the second iteration.
By contrast, if $a_1=1$, then $\rank(\openone -E_1)=d-1$, and the coin state at position $x=0$ is orthogonal to $|1\>$  after step 2(b) since the walker cannot move from $x=1$ to $x=0$;  at the beginning of the second iteration, we have $j=1$, and the coin state at position $x=0$ would be orthogonal to $|d-1\>$ due to the application of $C_{1,1}^{3}$, which interchanges the basis kets $|1\>$ and $|d-1\>$.

Next,  after $i=k-1$ iterations ($2\leq k\leq n-1$), suppose the quantum walk has generated POVM elements $E_1,E_2\dots,E_{i}$ at positions $x=2i,2i-2,\dots,2$, respectively, so that
the POVM element corresponding to position $x=0$ reads
\begin{equation}
\Gamma_{i}=\openone - \sum_{l=1}^{i} E_l=\sum_{l=i+1}^nE_l.
\end{equation}
For the convenience of the following discussion, we also define $\Gamma_0:=\openone$.
In the $k$th iteration, step 2(a) separates the element $\Gamma_{k-1}$ into $\widetilde{\Pi}_{k}=\tilde{a}_{k}|\psi_{k}\rangle\langle\psi_{k}|$ at position $x=1$ and $\Gamma_{k-1}-\tilde{a}_{k}|\psi_k\rangle\langle\psi_k|$ at position $x=-1$ and $0$.
Below we shall prove by induction the following three conclusions, which underpin the universality of the algorithm proposed above.
\begin{enumerate}
	\item $\tilde{a}_{i}=a^{\max}_i$, where $a_i^{\max}$ is the maximum value of any real number $a'$ that obeys $\Gamma_{i-1}-a' |\psi_i\>\<\psi_i|\geq0$; therefore,  an appropriate angle $\theta_i$ can always be chosen such that after step 2(b), the desired POVM element $E_i$ is generated at position $x=2$;
	\item $\rank(\Gamma_{i})=d-j$,
	where $j$ is the parameter at the end of the $i$th iteration, therefore, the inequality $j\leq d-1$ always holds in the running of the above algorithm given  that $\rank(\Gamma_{i})\geq \rank(E_{i})\geq 1$;
	\item if $j\geq1$ at the end of the $i$th iteration, then the coin state at position $x=0$ would be orthogonal to $|d-1\>, |d-2\>, \dots, |d-j\>$ for any initial state of the quantum walk.
\end{enumerate}
The three conclusions hold  when $i=1$. To achieve our goal,
we shall prove these conclusions for $i=k$, assuming that they hold for  $i\leq k-1$ with $2\leq k\leq n-1$.

By the induction hypothesis with $i=k-1$, we deduce that
\begin{equation}\label{eq:dim0}
\dim\big(\mathrm{span} \left\{|\phi_{k0}\rangle,|\phi_{k1}\rangle, \dots, |\phi_{k,d-1}\rangle\right\}\big)\leq d-j,
\end{equation}
where $j$ is the parameter at the beginning of the $k$th iteration or the end of the $(k-1)$th iteration.
In addition, we shall prove a stronger conclusion, namely,
\begin{equation}\label{eq:dim}
\dim\bigl(\mathrm{span} \left\{|\phi_{k1}\rangle, \dots, |\phi_{k,d-1}\rangle\right\}\bigr)<d-j.
\end{equation}
Suppose on the contrary that we have the equality
$\dim\big(\mathrm{span} \{|\phi_{k1}\rangle, \dots, |\phi_{k,d-1}\rangle\}\big) = d-j$,
then the ket $|\phi_{k0}\rangle$ can be expressed as a superposition of $|\phi_{k1}\rangle, \dots, |\phi_{k,d-1}\rangle$, say
\begin{equation}
|\phi_{k0}\rangle=b_1|\phi_{k1}\rangle+b_2|\phi_{k2}\rangle+\dots+b_{d-1}|\phi_{k,d-1}\rangle.
\end{equation}
Let  $|\delta\rangle$ be a normalized ket that is proportional to
\begin{equation}
|\delta\rangle\propto|\psi_{k}\rangle-\big( b_1|\psi_{k\perp1}\rangle+b_2|\psi_{k\perp2}\rangle+\dots+b_{d-1}|\psi_{k\perp d-1}\rangle \big). \end{equation}
If we choose $|\delta\rangle$ as the initial coin state of the quantum walk, then  the coin state at position $x=0$ at the beginning of the $k$th iteration would be
\begin{equation}
|\delta'\rangle\propto|\phi_{k0}\rangle-\big( b_1|\phi_{k1}\rangle+b_2|\phi_{k2}\rangle+\dots+b_{d-1}|\phi_{k,d-1}\rangle \big) =0.
\end{equation}
So the probability of finding the walker at position $x=0$ is zero, which contradicts the fact that $\<\delta|\Gamma_{k-1}|\delta\>\geq \<\delta|E_{k}|\delta\>>0$.
This contradiction confirms \eref{eq:dim}.

Owing to \eref{eq:dim}, we can choose the ket $|\eta_{k}\rangle$ such that it is  orthogonal to $|d-l\>$ for $l=1,2,\ldots,j$  in addition to $|\phi_{kl}\>$ for $l=1,2,\ldots, d-1$, and  set  $|\eta_{k\perp m}\rangle=|m\>$ for $m=d-1, d-2, \dots, d-j$, as mentioned in the algorithm. Then the  coin operator $C_{k,j}^{1}$ takes on  the  form
\begin{equation}
C_{k,j}^{1}=\begin{pmatrix} M_{d-j}&0\\0&{\openone_{j}} \end{pmatrix}.
\end{equation}
After applying the operator $C_{k,j}^{1}$, the coin state at position $x=0$ is still orthogonal to $|d-l\>$ with $l=1,2,\ldots,j$ for any initial state as is the case before the operation.
After the subsequent translation, the coin state at position $x=-1$ is proportional to $|1\>$, and the coin state at position $x=0$ is orthogonal to $|0\>, |1\>$ as well as $|d-1\>, |d-2\>, \dots, |d-j\>$.

Let $U$ be the unitary operator generated by the coin operators and translation operator, which governs  the evolution of the joint system composed of the coin and  walker. Then the  probability of finding the particle at position $x=-1$ can be written as
\begin{align}
p_{-1} &=\tr [U(|0\>\<0|\otimes |\varphi\>\<\varphi|)U^\dag (|-1\>\<-1|\otimes\openone )]\nonumber\\
&=\tr [U(|0\>\<0|\otimes |\varphi\>\<\varphi|)U^\dag (|-1\>\<-1|\otimes|1\>\<1| )]\nonumber\\
&=\tr(\Omega_{-1} |\varphi\>\<\varphi|),
\end{align}
where
\begin{equation}\label{eq:Omega1m}
\Omega_{-1}=\Tr_{\rmW}\{(|0\>\<0|\otimes\openone)U^{\dagger} \left(|-1,1\>\<-1,1|\right) U\}
\end{equation}
denotes the POVM element corresponding to position $x=-1$,
and "$\Tr_{\rmW}$" denotes the partial trace over the walker degree of freedom. Equation \eqref{eq:Omega1m} implies that  $\rank(\Omega_{-1})\leq 1$.
By the same token,
\begin{widetext}
	\begin{equation}\label{eq:Omega_{0}}
\Omega_{0}=\begin{cases}
\Tr_{\rmW}\{(|0\>\<0|\otimes\openone)U^{\dagger} [|0\>\<0|\otimes(|2\>\<2|+|3\>\<3|+\dots+|d-j-1\>\<d-j-1|)] U\} & j\leq d-3,\\
0 &j\geq d-2,
\end{cases}
\end{equation}
\end{widetext}
which implies the inequality
$\rank(\Omega_0)\leq d-j-2$ whenever $j\leq d-2$.

To proceed, we need to distinguish two cases depending on the value of $j$ at the beginning of the $k$th iteration.
\begin{enumerate}
	\item[Case A.] $j\leq d-2$\\
In this case, we have
\begin{align}
\ & \  d-j-1\leq \rank(\Gamma_{k-1})-\rank(\tilde{a}_{k}|\psi_{k}\rangle\langle\psi_{k}|)\nonumber\\
  &\leq \rank(\Gamma_{k-1}-\tilde{a}_{k}|\psi_{k}\rangle\langle\psi_{k}|)=\rank(\Omega_{-1}+\Omega_0)\nonumber\\
  &\leq \rank(\Omega_{-1})+\rank(\Omega_0)\leq d-j-1,
\end{align}
so that  $\rank(\Gamma_{k-1}-\tilde{a}_{k}|\psi_{k}\rangle\langle\psi_{k}|)=\rank(\Gamma_{k-1})-1$. By virtue of  \lref{lem:1} below we can deduce that $\tilde{a}_{k}=a^{\max}_k$.

If $\tilde{a}_{k}>a_k$ and $\cos\theta_k<1$, then the value of $j$ does not change during the $k$th iteration. After the iteration, the coin state at position $x=0$ is orthogonal to $|d-1\>, |d-2\>, \dots, |d-j\>$ for any initial state, and $\rank(\Gamma_k)=\rank(\Gamma_{k-1}-\tilde{a}_{k}\cos^2\theta_{k}|\psi_k\rangle\langle\psi_k|)=d-j$ according to  \lref{lem:1} below given that  $\tilde{a}_{k}\cos^2\theta_{k}<a^{\max}_k$.
	
By contrast, if  $\tilde{a}_{k}=a_k$ and $\cos\theta_k=1$, then the coin state at position $x=0$  after step 2(b) is orthogonal to $|1\>,|d-1\>, |d-2\>, \dots, |d-j\>$ for any initial state.
	In step 2(d), the value of $j$ increases by 1. After the $k$th iteration, the coin state at position $x=0$ is  orthogonal to $|d-1\>, |d-2\>, \dots, |d-j\>$ due to the application of $C_{k,j}^{3}$ (note that the value of $j$ has been updated), and we have $\rank(\Gamma_k)=\rank(\Gamma_{k-1}-\tilde{a}_{k}|\psi_{k}\rangle\langle\psi_{k}|)=d-j$  according to  \lref{lem:1} again.
	
\item[Case B.] $j=d-1$\\
In this case,
  the coin state at position $x=0$ at the beginning of the $k$th iteration is orthogonal to $|1\>, |2\>, \ldots, |d-1\>$ for any initial state by  the induction hypothesis; in other words, the state is proportional to $|0\>$. In addition,  $\rank(\Gamma_{k-1})=1$, which implies that $E_m\propto\Gamma_{k-1}=a^{\max}_k|\psi_k\rangle\langle\psi_k|$ for $m=k, k+1, \dots, n$, so that
\begin{equation}\label{eq:akmaxak}
a^{\max}_k=\sum_{m=k}^n a_m.
\end{equation}
Observing  that  $\<\psi_{k\perp m}|\Gamma_{k-1}|\psi_{k\perp m}\>=0$ for $m=1, 2, \dots, d-1$, we deduce that $|\phi_{km}\>=0$, which also follows from \eref{eq:dim}. Therefore, $|\eta_{k\perp m}\>=|m\>$ for $m=1, 2, \dots, d-1$, and $|\eta_k\>=|0\>$ up to an irrelevant phase factor, which implies that $C_{k,d-1}^{1}=\openone$.
After step 2(a), we have $\Omega_1=\Gamma_{k-1}$, so the equality $\tilde{a}_{k}=a^{\max}_k$ holds.
The coin state at position $x=0$ is proportional to $|1\>$
after step 2(b), but it is proportional to $|0\>$  after step 2(c) due to the application of $\NOT$. In addition,
 \eref{eq:akmaxak} implies that $\cos\theta_k=\sqrt{ a_k/ a^{\max}_k}<1$, so step 2(d) is irrelevant and  the value of $j$ does not change after $j$ reaches the value of $d-1$.
At the end of the iteration, $j=d-1$, $\rank(\Gamma_{k})=1=d-j$, and the coin state at position $x=0$ is orthogonal to $|d-1\>, |d-2\>, \dots, |d-j\>$, as expected.

\end{enumerate}

\subsection{An auxiliary lemma}
Here we prove an important lemma that underpins the proof of universality presented above.
\begin{lemma}\label{lem:1}
Suppose $A$ and $B$ are positive semidefinite operators  on a $d$-dimensional Hilbert space, $B$ has rank~1, and $\supp(B)\subset \supp(A)$. Let  $a$ be a complex number, then $\rank (A-aB)=\rank(A)-1$ if $a=a_{\max}$ and $\rank (A-aB)=\rank(A)$ otherwise, where  $a_{\max}$ is the maximum of any real number $a$ that obeys $A-aB\geq0$.
\end{lemma}
Here the rank condition $\rank (A-aB)=\rank(A)-1$ automatically guarantees that $a$ is a real number.

\begin{proof}
	Due to the assumption $\supp(B)\subset \supp(A)$, it suffices to consider this problem in $\supp(A)$,
	so we can assume that $A$ has full rank without loss of generality. Therefore,
	\begin{align}
	\rank (A-aB)&=\rank\big[ A^{-1/2}(A-aB)A^{-1/2}\big]\nonumber\\
                &=\rank (\openone-aA^{-1/2}BA^{-1/2}).
	\end{align}
    Given that $A^{-1/2}BA^{-1/2}$ has rank 1, it is clear that
	\begin{equation}
	a_{\max}=\bigl\|A^{-1/2}BA^{-1/2}\bigr\|^{-1}.
	\end{equation}
	In addition, we have	$\rank (A-aB)\geq \rank(A)-1$, and $A-aB$ is rank deficient iff $a=a_{\max}$, which confirms the lemma.

	Incidentally, the value of $a_{\max}$ is in general given by
	\begin{equation}
	a_{\max}=\bigl\|(A^+)^{1/2}B(A^+)^{1/2}\bigr\|^{-1},
	\end{equation}
	where $A^+$ is the Moore-Penrose generalized inverse of $A$.
\end{proof}

\end{document}